\documentclass[11pt]{article}

\usepackage{bbm}
\usepackage{graphicx}

\begin{document}

\title{Success criteria for quantum search on graphs}         
\author{Avatar Tulsi\\
        {\small Department of Physics, IIT Bombay, Mumbai - 400 076, India} \\ {\small tulsi9@gmail.com}}

\maketitle

\begin{abstract}
We consider quantum search on graphs. Recently, it has been shown that the graph properties like connectivity, global symmetry, or regularity cannot serve as a reliable criteria that must be satisfied by a graph to allow a successful quantum search. It is an open question whether it is possible to find such a criteria. We solve this question by giving an affirmative answer.  
\end{abstract}

\section{Introduction}

	Quantum search algorithm is one of the two most important quantum algorithms~\cite{grover1,grover2}. It started the extensive work on more general quantum search algorithms. Quantum search on graphs was introduced in ~\cite{childs1} where the search Hamiltonian uses the Laplacian of graph which couples only those vertices connected through graph edges. Many graphs are known to allow fast quantum search like the complete graphs, hypercubes, Paley graphs etc. At the same time, many graphs are known to deny fast quantum search like the 2- or 3-dimensional cubic periodic lattices, simplex of complete graphs etc. More than a decade has passed since its introduction but so far, there is no comprehensive theory of quantum search on graphs. Recently, several attempts have been made for this. It has been shown that contrary to previous intuitions, the properties of graph like connectivity, global symmetry, or regularity are \emph{not} a reliable criteria for fast quantum search on graphs~\cite{connectivity,globalsymmetry,regularity}. It has also been shown that \emph{almost all} randomly chosen graphs allow a fast quantum search~\cite{erdosrenyi}. 

	However, it remains an open question whether it is possible to find a reliable criteria which must be satisfied by a graph to allow a fast quantum search. In this paper, we solve this open question by giving an affirmative answer. We present an analysis of quantum search on graphs. The paper is organized as follows. In next Section, we present the analysis and then we discuss several examples to demonstrate our anlysis in Sections $3$ and $4$. Section 3 deals with the complete graph and related graphs while Section 4 deals with other important graphs like cubic periodic lattices and strongly regular graphs. We discusse and conclude in Section 5.	

\section{Quantum search on graphs} 

	We consider only undirected graphs with no self loops. Let $N$ be the total number of graph vertices labeled by $i \in \{1, 2,\ldots, N\}$. In quantum mechanics, these vertices are represented by the $N$ basis states $|i\rangle$ of an $N$-dimensional Hilbert space. A graph can be defined by its adjacency matrix $A$ whose diagonal elements $A_{ii}$ are zero for all $i$ and off-diagonal elements $A_{ij}$ are $1$ if and only if the vertices $i$ and $j$ are connected through an edge of a graph. If there is no graph edge connecting $i$ and $j$ then $A_{ij}$ is zero. The degree $D(i)$ of a vertex $i$ is the total number of vertices with which $i$ is connected through a graph edge. The degree matrix $D$ of a graph is a diagonal matrix whose elements are $D_{ii} = D(i)$. The Laplacian $L$ of a graph is given by $L = D-A$. 

	To do quantum search on graphs, we choose the initial state of our system to be $|s\rangle$ which is a uniform superposition of all vertices and also a ground state of the Laplacian. Our goal is to evolve it to a particular vertex $|t\rangle$ (the target state) which is a solution to a given search problem. We are provided an oracle which can easily identify $|t\rangle$ to implement the projector Hamiltonian $|t\rangle\langle t|$. Our strategy is to evolve our system under the following time-independent Hamiltonian 
\begin{equation} 
H = \gamma L - |t\rangle\langle t|,
\end{equation} 
where $\gamma$ is the jumping rate. Without the term $|t\rangle\langle t|$, the above Hamiltonian is simply $\gamma L$ which does not change the initial state $|s\rangle$ as it is a ground state of $L$. But, with this term, after an evolution time of  $\tau$, the state evolves to $|\psi(\tau)\rangle = e^{-\imath H\tau}|s\rangle$. We want to find the evolution time $T$ for which $|\psi(T)\rangle$ has maximum possible overlap with the target state $|t\rangle$. 

	To do so, we need to find the relevant eigenspectrum of $H$. The Laplacian $L$ is known to be a positive semidefinite matrix and its eigenspectrum can be written as 
\begin{equation} 
L|z\rangle = E_{z} |z\rangle, \ \ E_{0}=0 \leq E_{1}, \ldots, \leq E_{N-1}. \label{Laplacianeigen}
\end{equation} 
Let $|\lambda\rangle$ be an eigenstate of $H$ with the eigenvalue $\lambda$. Then we have 
\begin{equation}  
H|\lambda\rangle = \lambda |\lambda\rangle = \gamma L |\lambda\rangle-\langle t|\lambda\rangle |t\rangle. 
\end {equation} 
Left multiplication by $\langle z|$ gives 
\begin{equation}  
\lambda\langle z|\lambda\rangle = \gamma \langle z|L |\lambda\rangle - t_{z} \langle t|\lambda\rangle, \ \ t_{z} = \langle z|t\rangle. 
\end {equation} 
Here we have chosen the eigenstates $|z\rangle$ of $L$ such that $t_{z}$ are real. Rearranging the terms in above equation and using $\langle z |L = E_{z}\langle z|$, we get 
\begin{equation}  
\langle z|\lambda\rangle = \langle t|\lambda\rangle t_{z}(\gamma E_{z}-\lambda)^{-1}. \label{ztlambda}
\end {equation} 
Multiplying above equation by $t_{z} = \langle t|z\rangle $ and summing over $z$, we get 
\begin{equation}  
\langle t|\lambda\rangle = \sum_{z} \langle t|z\rangle \langle z|\lambda\rangle = \langle t|\lambda\rangle \sum_{z}t_{z}^{2}(\gamma E_{z}-\lambda)^{-1}
\end {equation} 
or 
\begin{equation}  
\sum_{z}t_{z}^{2}(\gamma E_{z}-\lambda)^{-1} = 1. \label{condition}
\end {equation} 
This is the condition that has to be satisfied by $\lambda$ to be an eigenvalue of $H$. In general, above equation is not so easy to solve. We show that for the typical cases of quantum search on graphs, we can make reasonable approximations to solve above equation.
	
	We assume that there exists an integer $m$ and an eigenvalue $\lambda$ such that $\gamma E_{m-1} \ll |\lambda| \ll \gamma E_{m}$. Precisely, for some $\chi \gg 1$, we assume
\begin{equation}
\gamma E_{m-1} < \frac{|\lambda|}{\chi},\ \ |\lambda|  < \frac{\gamma E_{m}}{\chi}\ . \label{assumption}
\end{equation}  
As $E_{z} \leq E_{m-1}$ for $z \leq m-1$ and $E_{z} \geq E_{m}$ for $z \geq m$, above assumption implies that all eigenvalues of the Laplacian are either very small or very large compared to $\lambda/\gamma$. With above assumption and little calculation, Eq. (\ref{condition}) can be rewritten as 
\begin{equation}
-\alpha^{2}\lambda^{-1}(1+\epsilon_{1}) + M_{1}\gamma^{-1} + \lambda  M_{2}(1+\epsilon_{2})\gamma^{-2} = 1. \label{equation2} 
\end{equation} 
Here $\epsilon_{1}$ and $\epsilon_{2}$ are functions of $\lambda$ whose magnitudes are upper bounded by $1/(\chi-1) \ll 1$. The quantities $\alpha$ and the moments $M_{r}$ are given by 
\begin{equation}
\alpha^{2} = \sum_{z < m}t_{z}^{2},\ \ \ M_{r} = \sum_{z \ geq m} t_{z}^{2}E_{z}^{-r}. \label{momentdefine} 
\end{equation} 
We define $\delta$ as 
\begin{equation}  
M_{1} = \gamma(1+\delta) \label{deltadefine} 
\end{equation} 
so that Eq. (\ref{equation2}) becomes 
\begin{equation} 
-\alpha^{2}\lambda^{-1}(1+\epsilon_{1}) + \delta + \lambda  M_{2}(1+\epsilon_{2})\gamma^{-2} = 0 \label{approximatequadratic} 
\end{equation} 
This is an approximate quadratic equation whose two solutions are $\lambda_{\pm}(1+O(1/\chi))$ where $\lambda_{\pm}$ are solutions of the following quadratic equation
\begin{equation} 
-\alpha^{2}\lambda^{-1} + \delta + \lambda  M_{2}\gamma^{-2} = 0. \label{quadratic} 
\end{equation} 
Their product $\lambda_{+}\lambda_{-}$ is $-\alpha^{2}\gamma^{2}/M_{2}$ so we can write 
\begin{equation} 
\lambda_{\pm} = \pm \frac{\alpha\gamma} {\sqrt{M_{2}}}(\tan \eta)^{\pm 1}. \label{lambdapm} 
\end{equation} 
The quantity $\eta$ is determined by the sum of roots $\lambda_{+} + \lambda_{-}$ which is $-\delta\gamma^{2}/M_{2}$. Using above equation, we get 
\begin{equation}  
\frac{\alpha\gamma} {\sqrt{M_{2}}}(\tan \eta-\cot\eta) = -\frac{\alpha\gamma} {\sqrt{M_{2}}}(2\cot 2\eta) = -\frac {\delta\gamma^{2}}{M_{2}}. \label{lambdapm2} 
\end{equation} 
So we have 
\begin{equation} 
\cot 2\eta = \frac{\delta\gamma}{2\alpha\sqrt{M_{2}}} =   \frac{\delta}{1+\delta} \frac{M_{1}}{ 2\alpha\sqrt{M_{2}} },\label{etaexpress} 
\end{equation} 
where we have used Eq. (\ref{deltadefine}). Eqs. (\ref{lambdapm}) and (\ref{etaexpress}) determine a pair of eigenvalues $\lambda_{\pm}$ of $H$ satisfying the assumption (\ref{assumption}). We show that these are the only relevant eigenvalues for our purpose as the evolution is mostly confined within the two-dimensional subspace spanned by the corresponding eigenstates $|\lambda_{\pm}\rangle$. 

	To find $|\lambda_{\pm}\rangle$, we choose them such that $\langle t|\lambda_{\pm}\rangle $ are real and positive. Using the normalization condition $\sum_{z} |\langle z|\lambda_{\pm}\rangle|^{2} = 1$ and Eq. (\ref{ztlambda}), we get 
\begin {equation} 
\frac{1}{\langle t|\lambda_{\pm}\rangle^{2}} = \frac{\alpha^{2}}{\lambda_{\pm}^{2}} + \frac{M_{2}}{\gamma^{2}} = \frac{M_{2}}{\gamma^{2}}\left[1+\tan^{\mp 2}\eta\right], \label{tlambdapmexpress}
\end {equation} 
where we have used Eq. (\ref{lambdapm}) and again ignored $O (1/\chi)$ terms. With little calculation, we get 
\begin{equation} 
\langle t |\lambda_{\pm}\rangle = \frac{\gamma}{\sqrt{M_{2}}}f_{\pm}(\eta),\ \ \ f_{\pm}(\eta) = \left(1+\tan^{\mp 2}\eta\right)^{-1/2}. \label{tlambdapmexpress2} 
\end {equation} 
It is easy to show that
\begin{equation}
f_{+}(\eta) = \sin \eta,\ \ f_{-}(\eta) = \cos \eta.
\end{equation}
As $\langle t|\lambda_{\pm}\rangle $ are chosen to be real and positive, the angle $\eta $ is chosen to satisfy $\eta \in [0, \pi/2] $. Putting above equation in Eq. (\ref{ztlambda}) and using the fact $|\lambda\rangle = \sum_{z}\langle z|\lambda\rangle|z\rangle$, we get 
\begin{equation}
|\lambda_{\pm}\rangle = \frac{\gamma}{\sqrt{M_{2}}}f_{\pm}(\eta) \sum_{z} \frac{t_{z}}{\gamma E_{z}-\lambda_{\pm}}|z\rangle. \label{lambdapmeigenstates}
\end{equation}
as expression for the eigenstates $|\lambda_{\pm}\rangle$. 

	We define the $|\sigma\rangle$ state as the normalised projection of the target state $|t\rangle$ on the $m$-dimensional subspace spanned by those eigenstates of $L$ whose eigenvalues satisfy $\gamma E_{z} \ll |\lambda_{\pm}|$. Using the assumption (\ref{assumption})and the definition of $\alpha$ in Eq. (\ref{momentdefine}), we write
\begin{equation}
|\sigma\rangle = (1/\alpha)\sum_{z<m}t_{z}|z\rangle, \label{sigmadefine}
\end{equation}
The importance of $|\sigma\rangle$ state becomes clear if we calculate $\langle \sigma|\lambda_{\pm}\rangle$. Using Eqs. (\ref{lambdapmeigenstates}) and (\ref{sigmadefine}) and the assumption (\ref{assumption}) that $\gamma E_{z} < |\lambda_{\pm}|/\chi$ for $z < m$, we find that 
\begin{equation}
\langle \sigma |\lambda_{\pm}\rangle = -\frac{\alpha \gamma}{\sqrt{M_{2}}}\frac{f_{\pm}(\eta)}{\lambda_{\pm}} = \mp f_{\mp}(\eta),
\end{equation}
where we have ignored $(O(1/\chi))$ terms and used Eq. (\ref{lambdapm}) for $\lambda_{\pm}$. We can write 
\begin {equation} 
|\sigma\rangle = -\cos \eta|\lambda_{+}\rangle + \sin \eta | \lambda_{-} \rangle.  \label{sigmaspan}
\end{equation} 
Thus the $|\sigma\rangle$ state is almost completely spanned by the two eigenstates $|\lambda_{\pm}\rangle$ of $H$. 

	After evolving the $|\sigma\rangle$ state under the Hamiltonian $ H $ for time $\tau $, we get the state 
\begin{equation} 
|\phi(\tau)\rangle = e^{-\imath H\tau}|\sigma\rangle =  -\cos \eta e^{-\imath (\lambda_{+}-\lambda_{-})\tau}|\lambda_{+}\rangle + \sin \eta | \lambda_{-} \rangle, 
\end{equation} 
where we have ignored the global phase factor. Using Eq. (\ref{tlambdapmexpress2}) and above equation,  we get 
\begin{equation} 
\langle t|\phi(\tau)\rangle = \frac{\gamma}{\sqrt {M_{2}}}\frac{\sin 2\eta}{2}\left( 1- e^{-\imath (\lambda_{+}-\lambda_{-})\tau}\right). 
\end{equation} 
Thus $ |\langle t|\phi(\tau)\rangle |$ is maximum for $ \tau = T $ given by 
\begin{equation} 
T = \frac {\pi}{\lambda_{+}- \lambda_{-}} = \frac{\pi \sqrt{M_{2}}}{2\gamma \alpha}\sin 2\eta. \label{evolutiontime}
\end{equation} 
The maximum value of $ \langle t|\phi(\tau)\rangle$ is 
\begin{equation} 
\langle t|\phi(T)\rangle = \frac{\gamma}{\sqrt {M_{2}}}\sin 2\eta = \frac{M_{1}}{\sqrt {M_{2}}}\frac{\sin 2\eta}{1+\delta}. \label{maximumoverlap} 
\end{equation} 
This is the maximum possible overlap with the target state during evolution and Eq. (\ref{evolutiontime}) determines the evolution time needed to obtain this maximum. 

\subsection{Special case: $\delta = 0$} 

We will mostly deal with the special case when $\delta$ is $0$ so that $\gamma $ is equal to its critical value $\gamma_{c}$, i.e. 
\begin{equation} 
\delta = 0 \Longrightarrow \gamma = \gamma_{c} = M_{1}. 
\end{equation} 
In this case, Eq. (\ref {etaexpress}) implies that $\eta$ is $\pi/4$ and $\tan\eta$ is $1$ so that Eq. (\ref {lambdapm}) implies that 
\begin{equation}
\delta = 0 \Longrightarrow \lambda_{\pm} = \pm \alpha M_{1}/\sqrt {M_{2}}. \label{balanceeigenvalues}
\end{equation} 
The assumption (\ref{assumption}) is then equivalent to
\begin{equation}
E_{m-1} \ll \alpha/\sqrt{M_{2}} \ll E_{m}, \label{balanceassumption} 
\end{equation}
for some value of $m$. As $\eta$ is $\pi/4$, $\sin 2\eta$ is $1$. Then Eqs. (\ref {evolutiontime} ) and (\ref {maximumoverlap}) imply that 
\begin {equation} 
\delta = 0 \Longrightarrow T= \frac {\pi \sqrt {M_{2}}}{2M_{1} \alpha}, \ \langle t|\phi (T)\rangle = \frac {M_{1}}{\sqrt {M_{2}}}. \label{optimalperformance}
\end {equation} 
Typically $\gamma $ must be very close to its critical value $\gamma_{c} $ for a successful quantum search. Otherwise $\delta$ is not close to zero making $|\cot 2\eta|$ very large so that either $\sin \eta $ or $\cos \eta $ is very close to $1$. Then $|\sigma\rangle $ is almost parallel to either $|\lambda_{+}\rangle $ or $|\lambda_{-}\rangle $ and it remains almost unchanged during evolution. Eq. (\ref{etaexpress}) and a little calculation implies that $|\cot 2\eta| \not\gg 1$ if and only if
\begin{equation}
|\gamma_{c}-\gamma| \not\gg 2\alpha \sqrt{M_{2}}.
\end{equation}
Typically $\alpha$ is very small and $M_{2}$ is $\Theta(1)$ hence the quantum search on graphs is extremely sensitive to the value of $\gamma$ near its critical value.

	In general, the $|\sigma\rangle$ state cannot be chosen as the initial state of evolution as it depends upon the $|t\rangle$ state. As mentioned earlier, the initial state is chosen to be $|s\rangle = \sum_{i}|i\rangle/\sqrt{N}$ which is independent of $|t\rangle$. It is easy to see that $|\sigma\rangle$ is $|s\rangle$ if and only if $|s\rangle$ is a non-degenerate ground state of $L$ and the non-zero eigenvalues of $L$ are much larger than $|\lambda_{\pm}|/\gamma$. We will mostly deal with the connected graphs for which the first condition is always satisfied and $\alpha$ is $\langle t|s\rangle = 1/\sqrt{N}$. In general, a graph can have $K$ connected components and the set $S$ of all graph vertices can be partitioned into $K$ mutually non-intersecting subsets $S_{k}$ ($k \in \{0, 1,\ldots, K-1\}$) where each subset is a set of all $N_{k}$ vertices of $k^{\rm th}$ component. Let $|s_{k}\rangle = \sqrt {1/N_{k}}\sum_{i \in S_{k}}|i\rangle$ denote the uniform superposition state of all vertices of the $k^{\rm th}$ component. Then it is easy to check that the states $|s_{k}\rangle$ form an orthonormal basis of the $K$-dimensional eigenspace of the Laplacian with eigenvalue $0$. Assuming that non-zero eigenvalues of the Laplacian are much larger than $|\lambda_{\pm}|/\gamma$, we find that $K$ is $m$. We choose $|z=k\rangle$ as $|s_{k}\rangle$ so that $E_{z}$ is $0$ for $z<m$. We denote the subset containing the target vertex $t$ as $S_{0}$. Then $|\sigma\rangle$ is $|s_{0}\rangle$ and $\alpha$ is $\langle t|s_{0}\rangle$ as $|t\rangle$ is orthogonal to other ground states $|z\neq 0\rangle$. Note that for a connected graph, the entire graph is a unique connected component hence $m$ is $1$ and $|s_{0}\rangle$ is $|s\rangle$ as expected.

\subsection{Summary of the analysis}

We summarize the main results of our analysis by presenting a step-by-step recipe to analyze quantum search on any graph. The steps are 
\\ 1. We find the eigenspectrum of the Laplacian $L$ of the graph and write it as
\begin{equation}
L|z\rangle = E_{z}|z\rangle,\ z \in \{0,1,\ldots,N-1\},\ \ E_{0} \leq E_{1} \leq \ldots \leq E_{N-1}.
\end{equation} 
We compute $t_{z}$ which is $\langle t|z\rangle$ and chosen to be real.
\\ 2. We intuitively choose a trial value of $m$ which determines $|\sigma\rangle$, $\alpha$, and the moments $M_{r}$ as 
\begin{equation}
|\sigma\rangle = \alpha^{-1}\sum_{z < m}t_{z}|z\rangle,\ \alpha = \langle t|\sigma\rangle = \sqrt{\sum_{z < m}t_{z}^{2}},\ \ M_{r} = \sum_{z\geq m}t_{z}^{2}E_{z}^{-r}
\end{equation}
Note that if $m$ is $1$ then $|\sigma\rangle$ is $|s\rangle$ and $\alpha$ is $1/\sqrt{N}$. In general, $|\sigma\rangle$ is the projection of $|t\rangle$ on the $m$-dimensional subspace spanned by the $|z < m\rangle$ eigenstates of $L$.
\\ 3. For our trial value of $m$, we check if our assumption
\begin{equation}
E_{m-1} \ll \alpha/\sqrt{M_{2}} \ll E_{m} \label{recipeassume}
\end{equation}
is correct. If not, we choose another trial value of $m$ and repeat above steps. If yes, we go to next step.
\\ 4. We evolve the initial state $|\sigma\rangle$ under the following Hamiltonian
\begin{equation}
H = M_{1}L - |t\rangle\langle t|,
\end{equation}     
where $\gamma$ is chosen to be its critical value $M_{1}$. There exists an optimal evolution time $T$ after which the evolved state $|\phi(T)\rangle$ is nearest to the desired target state $|t\rangle$. We have
\begin{equation}
T = (\pi \sqrt{M_{2}})/(2M_{1} \alpha), \ \ \langle t|\phi(T)\rangle = M_{1}/\sqrt{M_{2}}. \label{recipeperformance}
\end{equation} 
These steps describe the analysis of quantum search on any graph. To demonstrate our recipe, we discuss several graphs as examples. 

\section{Special Cases I: Complete graph and related graphs}

	We begin with the simplest example of a complete graph $C(N)$ of $N$ vertices, where all vertices are connected with each other through graph edges. Thus all off-diagonal elements of the adjacency matrix $A$ are $1$ and we have $A = J_{N} - \mathbbm{1}_{N}$ where $J_{N}$ and $\mathbbm{1}_{N}$ are the $N \times N$ all-ones matrix (whose all elements are $1$) and the identity matrix respectively. The degree $D(i)$ is $N-1$ for all vertices of $C(N)$ so that the degree matrix $D$ is $(N-1)\mathbbm{1}_{N}$. The Laplacian is
\begin{equation}
L[C(N)] = D-A = N\mathbbm{1}_{N} - J_{N} \label{CGlaplacian}
\end{equation} 
The two eigenvalues of $J_{N}$ are: $N$ for which $|s\rangle$ is the eigenstate and $0$ for which the $N-1$ dimensional eigenspace is orthogonal to $|s\rangle$. Thus the eigenspectrum of $L[C(N)]$ is defined by 
\begin{equation}
|0\rangle =|s\rangle,\ E_{0} = 0,\ E_{z\neq 0} =  N.
\end{equation}
We choose $m$ to be $1$ so $|\sigma\rangle$ is $|s\rangle$ and $\alpha$ is $1/\sqrt{N}$. Also $\sum_{z \geq 1} t_{z}^{2}$ is $1-(1/N)$ and as $E_{z \neq 0}$ is $N$, $M_{r}$ is $N^{-r}(1+O(1/N)) \approx N^{-r}$. Thus $\alpha/\sqrt{M_{2}}$ is $\sqrt{N}$ and the assumption (\ref{recipeassume}) is satisfied. The evolution Hamiltonian is chosen to be $L[C(N)]/N - |t\rangle\langle t|$ and Eq. (\ref{recipeperformance}) implies that the evolved state $|\phi(T)\rangle$ is exactly the desired target state $|t\rangle$ if the evolution time $T$ is chosen to be $\pi \sqrt{N}/2$. This is the best possible quantum search on a graph~\cite{optimal}.

	Next, we consider the joined complete graph $JC(N)$ of $N$ vertices. In ~\cite{connectivity}, this was presented as an example of a graph which allows a successful quantum search despite its low connectivity. To obtain a $JC(N)$, we join two complete graphs $C_{g}(N/2)$ for $g \in \{1,2\}$, of $N/2$ vertices each, by adding a joining edge that connects a vertex $a$ of $C_{1}(N/2)$ to a vertex $b$ of $C_{2}(N/2)$. With this definition, the Laplacian of $JC(N)$ can be written as 
\begin{equation}
L[JC(N)] = L_{B,2} + V_{JC},\ \ L_{B,2} = L[C_{1}(N/2)] \oplus L[C_{2}(N/2)], \label{JClaplacian1}
\end{equation}  
where $L_{B,2}$ is a block-diagonal matrix of two blocks with each block being a Laplacian $L[C(N/2)]$ of a complete graph. The matrix $V_{JC}$ corresponds to the joining edge and its all elements are zero except $4$ elements: $(V_{JC})_{aa}$ and $(V_{JC})_{bb}$ are $1$ whereas $(V_{JC})_{ab}$ and $(V_{JC})_{ba}$ are $-1$. We partition the set $S$ of $N$ vertices of $JC(N)$ into two complementary subsets $S_{g}$ ($g \in \{1,2\}$) of $N/2$ vertices where $S_{g}$ corresponds to the vertices of $C_{g}(N/2)$. As $L_{B,2}$ is a direct sum of $L[C(N/2)]$, Eq. (\ref{CGlaplacian}) can be used to write the eigenspectrum of $L_{B,2}$ as 
\begin{equation}
L_{B,2}|s_{g}\rangle = 0,\ \ |s_{g}\rangle = \sqrt{2/N}\sum_{i \in S_{g}}|i\rangle, \ \ g \in \{1,2\},\ \ L_{B,2}|\perp\rangle_{JC} = N/2, 
\end{equation}
where $|\perp\rangle_{JC}$ denotes any state within the $N-2$ dimensional eigenspace of $L_{B,2}$ which is orthogonal to both $|s_{1}\rangle$ and $|s_{2}\rangle$.

	We treat the extra term $V_{JC}$ as a small perturbation and use the perturbation theory to calculate the eigenspectrum of $L[JC(N)]$. The norm of $V_{JC}$ is $2$ but the energy difference between $|s_{g}\rangle$ states and $|\perp\rangle_{JC}$ states is $N/2$, so the mixing between these states due to $V_{JC}$ is of the order of $O(1/N)$ which can be ignored for large $N$. Also, the perturbed eigenvalues of $|\perp\rangle_{JC}$ will be within the interval $\{(N/2)\pm 2\}$. The double-degeneracy of $|s_{g}\rangle$ states gets splitted by $V_{JC}$ as $\langle s_{g}|V_{JC}|s_{g'}\rangle$ is $\pm 2/N$ where $+$ sign holds for $g = g'$ and $-$ sign holds for $g \neq g'$. With little calculation, we find the eigenspectrum of $L[JC(N)]$ to be
\begin{eqnarray}
|0\rangle = (1/\sqrt{2})(|s_{1}\rangle + |s_{2}\rangle), & E_{0} = 0 \\ \nonumber |1\rangle =  (1/\sqrt{2})(|s_{1}\rangle - |s_{2}\rangle), & E_{1} = 4/N \\ \nonumber |z>1\rangle = |\perp\rangle_{JC}, & E_{z>1} \in \{(N/2) \pm 2\}.
\end{eqnarray} 
We choose $m$ to be $2$. Then $|\sigma\rangle$ is the projection of $|t\rangle$ on the two-dimensional subspace spanned by the eigenstates $\frac{1}{\sqrt{2}}(|s_{1}\rangle \pm |s_{2}\rangle)$. Thus $|\sigma\rangle$ is either $|s_{1}\rangle$ or $|s_{2}\rangle$ depending upon whether $t$ is a vertex of $C_{1}(N/2)$ or $C_{2}(N/2)$. In both cases, $\alpha$ is $\sqrt{2/N}$ and $\langle \sigma |s\rangle$ is $1/\sqrt{2}$. Then $\sum_{z \geq 2}t_{z}^{2}$ is $1-(2/N)$ and $E_{z \geq 2} \in \{\frac{N}{2} \pm 2\}$ implies that $M_{r}$ is $(2/N)^{r}(1+O(1/N)) \approx (2/N)^{r}$. Thus $\alpha/\sqrt{M_{2}}$ is $\sqrt{N/2}$ and the assumption (\ref{recipeassume}) is satisfied. Evolving the $|\sigma\rangle$ state under the evolution Hamiltonian $(2/N)L[JC(N)]-|t\rangle\langle t|$ for time $T = \pi \sqrt{N/8}$ will yield the target state $|t\rangle$. If the initial state is $|s\rangle$ then, as $\langle \sigma|s\rangle$ is $1/\sqrt{2}$ and the evolution is a unitary transformation, we get a state having an overlap of $1/\sqrt{2}$ with $|t\rangle$ after evolution time $T$. A measurement will yield $t$ with a probability of $1/2$. This matches with the results of ~\cite{connectivity}. The total time complexity is $O(\sqrt{N})$ and a successful quantum search is possible on a joined complex graph. 

	We now consider the "simplex of complete graphs" of $N = R(R+1)$ vertices, $SC(N)$, which has been discussed in ~\cite{connectivity} as an example of a graph which does not allow an efficient quantum search despite its high connectivity. To get a $SC(N)$, we start with $R+1$ complete graphs of $R$ vertices, $C_{g}(R)$, for $g \in \{1,2,\ldots,R+1\}$. We then introduce connecting edges between all complete graphs which connects each vertex in a complete graph to a different complete graph. Each connecting edge joins a pair of two vertices of two different complete graphs and this pair is unique as no other connecting edge starts or ends at these two vertices. This is formally a first-order truncated $R$-simplex lattice ~\cite{simplexlattice} and it has $R^{2}(R+1)$ edges. We consider the more general weighted version of this graph where the edges within complete graphs have weight $1$, but edges between complete graphs (the connecting edges) have weight $w \ll \sqrt{R}$~\cite{weightedsimplex}.

	The Laplacian of $SC(N)$ can be written as
\begin{equation}
L[SC(N)] = L_{B,R+1} + V_{SC},\ \ L_{B,R+1} = \bigoplus_{g=1}^{R+1}L[C_{g}(R)],\ \ \|V_{SC}\| = w+1 \ll \sqrt{R}. \label{SClaplacian1}
\end{equation}  
Here $L_{B,R+1}$ is a block-diagonal matrix of $R+1$ blocks with each block being a Laplacian $L[C(R)]$ of a complete graph. The matrix $V_{SC}$ corresponds to the connecting edges. As there is exactly one connecting edge for each vertex, the diagonal elements of $V_{SC}$ are $1$. The off-diagonal elements $(V_{SC})_{ij}$ are $-w\delta_{j,j'(i)}$ where $\delta_{j,j'(i)}$ is the Kronecker's delta function and $j'(i)$ is the unique vertex joined with vertex $i$ through a connecting edge. By definition, $i \in C_{g}(R)$ and $j'(i) \in C_{g'}(R)$ imply that $g \neq g'$. Thus the norm of $V_{SC}$ is $w+1 \ll \sqrt{R}$. We partition the set $S$ of $N$ vertices of $SC(N)$ into mutually different subsets $S_{g}$ ($g \in \{1,2,\ldots,R+1\}$) of $R$ vertices where $S_{g}$ corresponds to the vertices of $C_{g}(R)$. As $L_{B,R+1}$ is a direct sum of $L[C(R)]$, Eq. (\ref{CGlaplacian}) can be used to write the eigenspectrum of $L_{B,R+1}$ as 
\begin{equation}
L_{B,R+1}|s_{g}\rangle = 0,\ \ |s_{g}\rangle = \sqrt{1/R}\sum_{i \in S_{g}}|i\rangle, \ \ g \in \{1,2,\ldots,R+1\},\ \ L_{B,R+1}|\perp\rangle_{SC1} = R, 
\end{equation}
where $|\perp\rangle_{SC1}$ denotes any state within the $N-R-1$ dimensional eigenspace of $L_{B,R+1}$ which is orthogonal to all $|s_{g}\rangle$.

	The extra term $V_{SC}$ is treated as a small perturbation while calculating the eigenspectrum of $L[JC(N)]$ using perturbation theory. Its norm is $w+1$ but the energy difference between $|s_{g}\rangle$ states and $|\perp\rangle_{SC1}$ states is $R$, so the mixing between these states due to $V_{SC}$ is of the order of $O(w/R)$ which can be ignored for large $R$ as $w \ll \sqrt{R}$. Also, the perturbed eigenvalues of $|\perp\rangle_{SC1}$ will be within the interval $\{R\pm w+1\} \approx R$ for large $R$. The $(R+1)$-degeneracy of $|s_{g}\rangle$ states gets splitted by $V_{SC}$ which is determined by the quantities $\langle s_{g}|V_{SC}|s_{g'}\rangle$. If $g=g'$, then the off-diagonal elements of $V_{SC}$ do not contribute to $\langle s_{g}|V_{SC}|s_{g}\rangle$ as they are non-zero only for two vertices corresponding to two different complete graphs, i.e. $g \neq g'$. The diagonal elements are $1$ and hence $\langle s_{g}|V_{SC}|s_{g}\rangle$ are $1$ for all $g$. Similarly, for $g \neq g'$, the diagonal elements do not contribute. Due to the off-diagonal elements, $|s_{g}\rangle = (1/\sqrt{R})\sum_{i \in S_{g}}|i\rangle$ gets transformed to the state $(-w/\sqrt{R})\sum_{i \in S_{g}}|j'(i)\rangle$. By definition, a given $g'$ contains only one out of $R$ vertices $j'(i)$ for all $i \in S_{g}$ and the overlap of $|s_{g'}\rangle$ state with that vertex is $1/\sqrt{R}$. Thus $\langle s_{g}|V_{SC}|s_{g'}\rangle$ is $-w/R$ for $g \neq g'$. With this, we find that in the $(R+1)$-dimensional basis orthonormally spanned by the states $|s_{g}\rangle$ for all $g$, the matrix $V_{SC}$ is equivalent to 
\begin{equation}
V_{JC} \equiv [1+(w/R)]\mathbbm{1}_{R+1} - (w/R)J_{R+1},
\end{equation}
where $\mathbbm{1}_{R+1}$ and $J_{R+1}$ are the identity and all-ones matrices respectively as defined earlier. There are two eigenvalues of $J_{R+1}$. First is $R+1$ for which the uniform superposition of all $|s_{g}\rangle$ states is the eigenstate. This  is easy to check that this uniform superposition is nothing but the $|s\rangle = (1/\sqrt{N})\sum_{i}|i\rangle$ state which is a uniform superposition of all $N$ vertices of the graph. Second eigenvalue is $0$ for which the eigenstate is any state $|\perp\rangle_{SC2}$ orthogonal to $|s\rangle$ but completely within the $(R+1)$-dimensional subspace orthogonally spanned by $|s_{g}\rangle$ states. 

	With these facts, we can write the eigenspectrum of $L[SC(N)]$ as
\begin{eqnarray}
|0\rangle = |s\rangle, & E_{0} = 1-w \\ \nonumber |1 \leq z\leq R\rangle =  |\perp\rangle_{SC2}, & E_{1\leq z \leq R} = 1+(w/R) \\ \nonumber |z>R\rangle = |\perp\rangle_{SC1}, & E_{z>R} \in \{R \pm w\}.
\end{eqnarray} 
Note that $E_{0}$ is not zero and to apply our analysis, we add a constant energy of $w-1$ to the Laplacian. This does not change the dynamics as it causes only an ignorable global phase factor. Assuming $1 \ll w \ll \sqrt{R}$, we can then write the approximate eigenspectrum of shifted Laplacian as 
\begin{eqnarray}
|0\rangle = |s\rangle, & E_{0} = 0 \\ \nonumber |1 \leq z\leq R\rangle =  |\perp\rangle_{SC2}, & E_{1\leq z \leq R} \approx w \\ \nonumber |z>R\rangle = |\perp\rangle_{SC1}, & E_{z>R} \approx R. \label{simplexcompleteeigen}
\end{eqnarray} 
The second order perturbation does not significantly change above eigenvalues as this change is upper bounded by $\|V\|^{2}/(E_{z>R}-E_{z \leq R})$ which is approximately $w^{2}/R \ll 1$ as $w \ll \sqrt{R}$. Let us denote the complete graph containing the target vertex $t$ as $C_{1}(R)$. Then the projection of $|t\rangle$ on $(R+1)$-dimensional subspace spanned by all $|s_{g}\rangle$ states is $|s_{1}\rangle$ and as $\langle t|s_{1}\rangle$ is $1/\sqrt{R}$, we find that, for large $R$,
\begin{equation}
t_{0} = \frac{1}{\sqrt{N}},\ \sum_{z = 1}^{R}t_{z}^{2} = \frac{1}{R} - \frac{1}{N} \approx \frac{1}{R},\ \sum_{z > R}t_{z}^{2} = 1-\frac{1}{R}+\frac{1}{N} \approx 1. \label{simplexcompletetz}
\end{equation}

	First, we choose $m$ to be $1$ so that $|\sigma\rangle$ is $|s\rangle$ and $\alpha$ is $1/\sqrt{N}$. Eqs. (\ref{simplexcompleteeigen}) and (\ref{simplexcompletetz}) imply that the moments $M_{r}$ are given by 
\begin{equation}
M_{r} = \frac{1}{Rw^{r}} + \frac{1}{R^{r}}. 
\end{equation}
Thus $M_{2} \approx 1/Rw^{2}$ as $w \ll \sqrt{R}$ and $\alpha/\sqrt{M_{2}}$ is $w\sqrt{R/N} \ll w$ for large $R$ and hence the assumption (\ref{recipeassume}) is satisfied. We evolve the initial state $|\sigma\rangle$ under the following Hamiltonian
\begin{equation}
H = M_{1}L - |t\rangle\langle t|,\ \  M_{1} = \frac{1}{R}\left(1+\frac{1}{w}\right).
\end{equation}     
After the optimal evolution time $T$, we get the state $|\phi(T)\rangle$ given by
\begin{equation}
T = \frac{\pi}{2}\frac{\sqrt{RN}}{w+1} = O\left(\frac{N^{3/4}}{w}\right), \ \ \langle t|\phi(T)\rangle = \frac{w+1}{\sqrt{R}}. \label{firststagetime}
\end{equation}
Note that $|\phi(T)\rangle$ has a considerable overlap with the target stae $|t\rangle$ if and only if $w^{2}$ is comparable to $R$. If $w^{2} \ll R$ then $\langle \phi(T)|t\rangle \ll 1$. Particularly, for an unweighted graph, $w$ is $1$ and $\langle \phi(T)|t\rangle$ is $2/\sqrt{R}$. 

	We show that despite this, the state $|\phi(T)\rangle$ contains sufficient information about the target state $|t\rangle$ as $|\phi(T)\rangle$ is very close to $|s_{1}\rangle$, the uniform superposition of $R$ vertices of the complete graph $C_{1}(R)$ containing the target vertex. To show this, we note that we have chosen $\gamma$ to be its critical value $M_{1}$ so $\eta$ is $\pi/4$ and $f_{\pm}(\eta)$ is $1/\sqrt{2}$. As $|\sigma\rangle$ is $|s\rangle$ when $m$ is $1$, Eq. (\ref{sigmaspan}) implies that $|s\rangle = (1/\sqrt{2})|\lambda_{-}\rangle -|\lambda_{+}\rangle$. The evolution changes the relative sign between two eigenstates $|\lambda_{\pm}\rangle$ and after an evolution time of $T$, we get the state $|\phi(T)\rangle = (1/\sqrt{2})(|\lambda_{+}\rangle + |\lambda_{-}\rangle)$ which has an overlap of $(w+1)/\sqrt{R}$ with the target state $|t\rangle$. Putting $f_{\pm}(\eta) = 1/\sqrt{2}$ and using the approximation $\gamma E_{0} \ll |\lambda_{\pm}| \ll \gamma E_{1}$, Eq. (\ref{lambdapmeigenstates}) can be rewritten as
\begin{equation}
|\lambda_{\pm}\rangle \approx \frac{1}{\sqrt{2}}\frac{\gamma}{\sqrt{M_{2}}}\left(\mp \frac{t_{0}}{\lambda_{+}}|0\rangle+ \sum_{z \neq 0}\frac{t_{z}}{\gamma E_{z}}|z\rangle\right),
\end{equation}   
where we have used Eq. (\ref{balanceeigenvalues}) which implies that $\lambda_{\pm} = \pm \lambda_{+}$ when $\gamma$ is $M_{1}$. Then we have
\begin{equation} 
|\phi(T)\rangle = \frac{1}{\sqrt{2}}(|\lambda_{+}\rangle + |\lambda_{-}\rangle) = \frac{1}{\sqrt{M_{2}}}\sum_{z \neq 0}\frac{t_{z}}{E_{z}}|z\rangle
\end{equation}
In case of the graph $SC(N)$, using Eq. (\ref{simplexcompleteeigen}) and $M_{2} \approx 1/Rw^{2}$, we get
\begin{equation}
|\phi(T)\rangle \approx w\sqrt{R}\left(\frac{1}{w}\sum_{z=1}^{R}t_{z}|z\rangle + \frac{1}{R}\sum_{z > R}t_{z}|z\rangle\right) 
\end{equation} 
Using Eq. (\ref{simplexcompletetz}), we find the lengths of the first and second terms within the bracket of the above equation as $w^{-1}\sqrt{\sum_{z=1}^{R}t_{z}^{2}} = (w\sqrt{R})^{-1}$ and $R^{-1}\sqrt{\sum_{z>R}t_{z}^{2}} = R^{-1}$ respectively. Thus the second term can be neglected as $w \ll \sqrt{R}$ and we get 
\begin{equation}
|\phi(T)\rangle \approx \sqrt{R}\sum_{z=1}^{R}t_{z}|z\rangle.  
\end{equation}
This is a normalized state which does not change with the addition of the state $\sqrt{R}t_{0}|0\rangle$ whose length is $\sqrt{R/N}$ (as $t_{0}$ is $1/\sqrt{N}$) which is negligible for large $N$. Thus $|\phi(T)\rangle$ is approximately $\sqrt{R}$ times the projection of $|t\rangle$ on the $(R+1)$-dimensional subspace spanned by the states $|s_{g}\rangle$ for all $g$. This projection is nothing but $(1/\sqrt{R})|s_{1}\rangle$ as the labels of the complete graphs are chosen such that $|t\rangle$ is an element of the first complete graph $C_{1}(R)$. Thus $|\phi(T)\rangle$ is $|s_{1}\rangle$ and its measurement will let us know the complete graph of $R$ vertices containing the target state. Our search becomes easier then as it reduces from searching a $SC(N)$ graph of $N$ vertices to searching a complete graph of $R$ vertices. 

	Searching a complete graph has been described earlier but that was done using the Laplacian of a complete graph. The question is whether we can efficiently search a complete graph using the Laplacian of $SC(N)$. The answer is yes and to show this, we choose $m$ to be $R+1$ rather than $1$. Then $|\sigma\rangle$ is the normalised projection of $|t\rangle$ on $(R+1)$-dimensional subspace spanned by $|s_{g}\rangle$ states for all $g$. Thus $|\sigma\rangle$ is $|s_{1}\rangle$ and $\alpha$ is $\langle t|s_{1}\rangle = 1/\sqrt{R}$. Eqs. (\ref{simplexcompleteeigen}) and (\ref{simplexcompletetz}) imply that the moments $M_{r}$ are given by 
\begin{equation}
M_{r} = \frac{1}{E_{z>R}^{r}}\sum_{z > R}t_{z}^{2} = R^{-r},
\end{equation}
so $M_{1} = 1/R$. This value of $M_{1}$ is different from the value $(1/R)(1+w^{-1})$ when $m$ was chosen to be $1$. With this value of $M_{1}$ in the evolution Hamiltonian and starting the evolution with the state $|\sigma\rangle = |s_{1}\rangle$, we find that $|\phi(T)\rangle$ is $|t\rangle$ as $\langle t|\phi(T)\rangle = M_{1}/\sqrt{M_{2}} = 1$. The evolution time $T$ is $\pi\sqrt{M_{2}}/(2M_{1} \alpha)$ which is $O(\sqrt{R})$ as $\alpha$ is $1/\sqrt{R}$.

	Thus, for a $SC(N)$, we basically do a two-stage quantum search as first demonstrated in ~\cite{connectivity,weightedsimplex}. In the first phase we choose $\gamma$ to be $R^{-1}(1+w^{-1})$ and evolve the initial state $|s\rangle$ to $|s_{1}\rangle$. This takes an evolution time of $O(N^{3/4}/w)$ according to Eq. (\ref{firststagetime}). In the second stage, we choose $\gamma$ to be $R^{-1}$ and evolve $|s_{1}\rangle$ to $|t\rangle$ in the time $O(\sqrt{R})$ which is $O(N^{1/4})$ as $N \approx R^{2}$ for large $R$. As $w \ll \sqrt{R}$, the total time complexity is $O(N^{3/4}/w)$ and we get optimal quantum search by choosing $w$ to be $\Theta(\sqrt{R})$. These results completely match with the results of ~\cite{connectivity,weightedsimplex}.

\section{Special Cases II: Other important graphs}

\subsection{Hypercube}

	For a $n$-dimensional hypercube, there are $N = 2^{n}$ vertices of graph. Each vertex $i$ is labeled by a $n$-bit binary string $x(i)$ whose $h^{\rm th}$ bit is denoted by $x_{h}(i)$ ($h\in \{1,2,\ldots,n\}$). Two vertices $i$ and $j$ are connected through a graph edge if and only if they differ in a single bit, i.e., the Hamming distance between $x(i)$ and $x(j)$ is $1$. Thus the degree of each vertex is $n$ and the degree matrix $D$ is $n\mathbbm{1}_{N}$. The adjacency matrix $A$ can be written as $\sum_{h = 1}^{n}A_{h}$ where $(A_{h})_{ij}$ are zero except when $i$ and $j$ differ only in the value of $h^{\rm th}$ bit in which case $(A_{h})_{ij}$ is $1$. Within the two-dimensional subspace spanned by the two vertices differing only in the value of $h^{\rm th}$ bit, the matrix $A_{h}$ is a $2 \times 2$ matrix whose off-diagonal elements are $1$ but the diagonal elements are $0$. Its eigenvalues are $\pm 1$ and the corresponding eigenstates are $(1/\sqrt{2})(|i\rangle \pm |j\rangle)$ where $i$ and $j$ differ only in the value of $h^{\rm th}$ bit. Representing these vertices by the $n$-bit strings, these eigenstates can be written as 
\begin{equation}
(1/\sqrt{2})(|x_{h \neq h'}(i,j)\rangle) \otimes (|0_{h}\rangle \pm |1_{h}\rangle)
\end{equation} 
where $|0_{h}\rangle$ or $|1_{h}\rangle$ is the $h^{\rm th}$ bit value of the vertex and $x_{h\neq h'}(i,j)$ represents the common values of the $n-1$ remaining bits $h' \neq h$ of the vertices $i$ and $j$. 

	As $A$  is $\sum_{h=1}^{n}A_{h}$ and each $A_{h}$ acts on a different bit, it is easy to check that the eigenstates of $A$ can be written in the form
\begin{equation}
\frac{1}{2^{n/2}}\left(\bigotimes_{h =1}^{p} \left(|0_{h}\rangle - |1_{h}\rangle \right) \otimes \bigotimes_{h =p+1}^{n} \left(|0_{h}\rangle + |1_{h}\rangle \right) \right),\ \ p\in\{0,1,\ldots,n\} 
\end{equation}  
and the corresponding eigenvalues are $(n-p)-p = n-2p$. The total number of the eigenstates for a fixed value of $p$ is the number in which first set of $p$ indices can be chosen out of $n$ possible indices of $h$ which is $^{n}C_{p}$. These are also the eigenstates $|z\rangle$ of the Laplacian $L$ with the eigenvalues $2p$ as $L$ is $D-A$ and $D$ is $n \mathbbm{1}_{N}$. Thus the non-degenerate ground state of the Laplacian, corresponding to $p=0$, is $|s\rangle =\sum_{i}|i\rangle/\sqrt{N}$ as expected. 

	To apply our analysis, we choose $m$ to be $1$ so that the initial state $|\sigma\rangle$ is $|s\rangle$ and $\alpha$ is $1/\sqrt{N}$. The moments $M_{r}$ can be rewritten as
\begin{equation}
M_{r} = \sum_{z \neq 0}\frac{t_{z}^{2}}{E_{z}^{r}} = \sum_{p \neq 0}\frac{t_{p}^{2}}{(2p)^{r}},\ \ t_{p}^{2} = \sum_{z,E_{z} = 2p}t_{z}^{2}.
\end{equation}  
The overlap of the target state $|t\rangle$ (which represents a unique vertex) with each eigenstate $|z\rangle$ is $\pm 1/\sqrt{N}$ and hence $t_{z}^{2}$ is $1/N$ for all $z$. As there are $^{n}C_{p}$ eigenstates with the eigenvalue $E_{z} = 2p$, we find $t_{p}^{2}$ to be $^{n}C_{p}/N$ so that
\begin{equation}
M_{r} = \frac{1}{N}\sum_{p \neq 0}\ \ ^{n}C_{p}\frac{1}{(2p)^{r}}.
\end{equation} 
The quantity $^{n}C_{p}$ is maximum at $p=n/2$ and decreases exponentially away from this, being relatively non-negligible only when $p \in \{\frac{n}{2}\pm O(\sqrt{n})\}$. Hence $M_{r} \approx 1/n^{r}$. So $\alpha/\sqrt{M_{2}}$ is $n/\sqrt{N}$ satisfying the assumption $E_{0} \ll \alpha/\sqrt{M_{2}} \ll E_{1}$ for large $N$ as $E_{0}$ is $0$ and $E_{1}$ is $2$. We choose the evolution Hamiltonian to be $H = L/n - |t\rangle\langle t|$ and start with the $|s\rangle$ state. After the optimal evolution time $T = O(\sqrt{N})$, we get the $|\phi(T)\rangle$ state which is the target state $|t\rangle$ as $\langle t|\phi(T)\rangle$ is $M_{1}/\sqrt{M_{2}} \approx 1$. 

	Our analysis gives an alternative to the analysis of the search on hypercube presented in ~\cite{hypercube1} and Appendix B of ~\cite{hypercube2}.

\subsection{Cubic Lattices}

	We now consider the case of a $d$ dimensional cubic periodic lattices where $d$ is fixed independent of the number of vertices $N$. This was first analysed in ~\cite{childs1}. Each vertex of the lattice can be represented by a $d$-component vector with components $x_{j} \in \{0,1,\ldots, N^{1/d}-1\}$. (The notation used in this subsection has different meanings as defined here.) The lattice is periodic in each direction with period $N^{1/d}$. The eigenstates of the Laplacian are $|z\rangle = |\phi(k)\rangle$ given by
\begin{equation}
\phi(k)\rangle = \frac{1}{\sqrt{N}}\sum_{x}e^{\imath k\cdot x }|x\rangle \Longrightarrow |\langle t|z\rangle| = 1/\sqrt{N},
\end{equation}    
where
\begin{equation}
k_{j} = \frac{2\pi m_{j}}{N^{1/d}},\ \ m_{j} \in \{0, \pm 1,\ldots, \pm \frac{1}{2}(N^{1/d}-1)\}.
\end{equation}
Without loss of generality, we have assumed $N^{1/d}$ to be odd. The corresponding eigenvalues are
\begin{equation}
E(k) = 2\left(d - \sum_{j=1}^{d}\cos (k_{j})\right).
\end{equation}
For small values of $k$, we have
\begin{equation}
E(k) \approx k^{2} = \frac{(2\pi m)^{2}}{N^{2/d}},\ \ k^{2} = k_{1}^{2} + \cdots +k_{d}^{2}. \label{cubicsmallk}
\end{equation}

	With $t_{z}^{2} = 1/N$, we get
\begin{equation}
M_{r} = \frac{1}{N}\sum_{k \neq 0}\frac{1}{[E(k)]^{r}}. 
\end{equation}
This is same as the quantity $S_{j,d}$ defined in Eq. (33) of ~\cite{childs1}. It has been discussed in detail there. For $d > 2r$, this can be approximated by an integral which converges to a constant value which is $\Theta(1)$. For $d = 2r$, there is a logarithmic divergence of this integral and then $M_{r}$ is $\Theta(\ln N)$. 

	As $r \in \{1,2\}$, for $d \geq 5$, $M_{r}$ is always $\Theta(1)$. Also, Eq. (\ref{cubicsmallk}) implies that the smallest non-zero eigenvalue for $d \geq 5$ is $\Theta(N^{-2/d}) \gg 1/\sqrt{N}$ for large $N$. Hence the assumption $E_{m-1} \ll \alpha/\sqrt{M_{2}} \ll E_{m}$ is satisfied for $m = 1$ as $\alpha$ is $1/\sqrt{N}$. As $M_{1}/\sqrt{M_{2}}$ is $\Theta(1)$, we get a state having a constant overlap with the target state after evolving for time $\Theta(\sqrt{N})$ which is the optimal performance.

	When $d$ is $4$ then $M_{1}$ is $\Theta(1)$ as $d > 2r$ for $r =1$ but $M_{2}$ is $\Theta(\log N)$ as $d =2r$ for $r =2$. Thus $\alpha /\sqrt{M_{2}}$ is $1/\Theta(\sqrt{N} \log N)$. Eq. (\ref{cubicsmallk}) implies that the smallest non-zero eigenvalue is $1/\sqrt{N}$ and hence the assumption $E_{m-1} \ll \alpha/\sqrt{M_{2}} \ll E_{m}$ is satisfied when $m$ is chosen to be $1$. As $M_{1}/\sqrt{M_{2}}$ is $1/\Theta(\sqrt{\log N})$, after the evolution time of $\Theta(\sqrt{N\log N})$, we get a state $|\phi (T)\rangle$ having an overlap of $1/\Theta(\sqrt{\log N})$ with the target state $|t\rangle$. Thus the quantum search is logarithmically slow compared to its optimal performance of $\Theta(\sqrt{N})$.

	When $d$ is $3$ then $M_{1}$ is $\Theta(1)$ as $d > 2r$ when $r$ is $1$. But when $r$ is $2$ then $d < 2r$ and in that case, small values of $k$ have a dominating contribution to $M_{2}$. Then $M_{2}$ is basically $S_{2,3}$ defined in Eqs. (38-40) of ~\cite{childs1} which implies that 
\begin{equation}
M_{2}  = 0.0265 N^{1/3} \Longrightarrow \alpha /\sqrt{M_{2}} = 6.143 N^{-2/3}.
\end{equation}
Also, Eq. (\ref{cubicsmallk}) implies that the smallest eigenvalue is $4\pi^{2}N^{-2/3} = 39.48 N^{-2/3}$. Hence the assumption $E_{m-1} \ll \alpha/\sqrt{M_{2}} \ll E_{m}$ is satisfied for the choice $m =1$. The quantity $M_{1}/\sqrt{M_{2}}$ is $\Theta (N^{1/3})$ and hence $|\phi(T)\rangle$ has a negligible overlap with the target state $|t\rangle$ for large $N$. So a successful quantum search is not possible in this case.

	Similar considerations hold when $d$ is $2$ except that then $M_{1}$ is $\Theta (\log N)$ rather than $\Theta(1)$ as $d = 2r$ for $r =1$. However, a successful quantum search is not allowed in this case also.

\subsection{Random Erd\"{o}s-Renyi graphs}

	A random Erd\"{o}s-Renyi graph $ER(N,P)$ of $N$ vertices is a graph where an edge between any two vertices exists with probability $P$ independently of all other edges. This random graph model was introduced by Erd\"{o}s and Renyi in their seminal work~\cite{erdos,renyi} and they studied the probability of a random graph to possess a certain property $Q$ like connectedness, presence of certain subgraphs, etc. They introduced the terminology stating that \emph{almost all graphs} have a property $Q$ if the probability of a random graph $ER(N,P)$ having the property $Q$ goes to $1$ in the asymptotic limit $N \rightarrow \infty$. For many properties $Q$, there exists a critical probability $P=P_{c}$ such that for $P > P_{c}$, almost all graphs have the property $Q$ but for $P < P_{c}$, almost all graphs do not have this property. For example, $P_{c}$ is the percolation threshold $\log(N)/N$ for the property of connectedness which implies that the graph is almost surely connected for $P > \log(N)/N$ but has almost surely isolated nodes for $P < \log(N)/N$. 

	In ~\cite{erdosrenyi}, the authors have studied the property of the optimality (i.e. a running time of $O(\sqrt{N})$ of quantum search on $ER(N,P)$ and found the critical value $P_{c}$ for this. They have chosen the evolution Hamiltonian to be $-\gamma A - |t\rangle \langle t|$ ($A$ is the adjacency matrix) rather than $\gamma L - |t\rangle \langle t|$ chosen here. Our analysis can be used by replacing $L$ by $-A$. Thus we choose $L$ to be $-A$ rather than $D-A$. Let the eigenspectrum of the adjacency matrix $A$ be given by
\begin{equation}
A|y\rangle = a_{y}|y\rangle,\ y\in\{0,1,\ldots,N-1\},\ \ a_{0} \geq a_{1} \geq  \ldots \geq a_{N-1}.
\end{equation}
Then the lowest eigenvalue of $L$ is $-a_{0}$ where $a_{0}$ is the highest eigenvalue of $A$. To use our analysis, we want this lowest eigenvalue to be $0$ so we add a constant energy term $a_{0}\mathbbm{1}_{N}$ to the Laplacian. Doing so does not change the dynamics and the new Laplacian is $a_{0}\mathbbm{1}_{N} - A$ whose eigenspectrum is given by 
\begin{equation}
L|z\rangle = E_{z}|z\rangle,\ |z\rangle = |y\rangle,\ E_{z} = a_{0}-a_{z} = a_{0}[1-(a_{z}/a_{0})].
\end{equation}
We choose $m$ to be $1$ in our analysis so the initial state $|\sigma\rangle$ is $|0\rangle$, the eigenstate of $A$ with maximum eigenvalue. As shown in Section III of the Supplemental Material of ~\cite{erdosrenyi}, the state $|0\rangle$ is very close to $|s\rangle = \sum_{i}|i\rangle/\sqrt{N}$ for almost all graphs $ER(N,P)$ as long as 
\begin{equation}
P > P_{c} = \log^{3/2}(N)/N.
\end{equation} 
Thus $\alpha$ is $1/\sqrt{N}$. In ~\cite{erdoseigen}, it is also shown that for $P \geq \log^{4/3}(N)/N$ (which is true as long as above inequality is true), the quantity $|a_{z \neq 0}/a_{0}|$ is upper bounded by a positive constant $c < 1$ for almost all graphs $ER(N,P)$. Then we have $a_{0}(1-c) \leq E_{z \neq 0} \leq a_{0}(1+c)$ and  as $\sum_{z \neq 0}t_{z}^{2} \leq 1$, the moments $M_{r}$ satisfy 
\begin{equation}
\frac{1}{a_{0}^{r}}\frac{1}{(1+c)^{r}} \leq M_{r} = \sum_{z \neq 0}\frac{t_{z}^{2}}{E_{z}^{r}} \leq \frac{1}{a_{0}^{r}}\frac{1}{(1-c)^{r}}  
\end{equation} 
Thus $\sqrt{M_{2}} \geq O(1/a_{0})$ so that $\alpha/\sqrt{M_{2}}$ is $O(a_{0}/\sqrt{N})$ and the assumption $E_{0} \ll \alpha/\sqrt{M_{2}} \ll E_{1}$ is satisfied for large $N$ as $E_{0}$ is $0$ and $E_{1}$ is $\Theta(a_{0})$. We choose the evolution Hamiltonian to be $H = M_{1}L - |t\rangle\langle t|$ where $M_{1}$ is $\Theta(1/a_{0})$. As $M_{1}/\sqrt{M_{2}}$ is always greater than $(1-c)/(1+c)$, the state $|\phi(T)\rangle$ satisfies $\langle t|\phi(T)\rangle \geq (1-c)/(1+c)$ and the evolution time $T$ is $\Theta(\sqrt{N})$. Thus the quantum search is optimal.

	\textbf{Complete graphs with missing edges:} The Erd\"{o}s-Renyi random graph $ER(N,P)$ can also be obtained from the complete graph by randomly deleting edges with probability $1-P$. Thus, the quantum search on complete graphs is inherently robust to random loss of edges. This also explains the success of quantum search on complete graphs with broken links (equivalent to missing edges) as first demonstrated in ~\cite{regularity}.	

	\textbf{Random regular graphs:} Our analysis gives an alternative proof of the Lemma $1$ of ~\cite{erdosrenyi} which introduces the assumption $|a_{z \neq 0}/a_{0}| \leq c < 1$. As discussed in Section II of the supplementary material of ~\cite{erdosrenyi}, this assumption is also true for a random graph sampled uniformly from the set of all regular graphs of
degree $d$ for $d \geq 3$. Thus the quantum search is also optimal for random regular graphs.

\subsection{Strongly regular graphs} 

A strongly regular graph of $N$ vertices, $SR(N)$, is defined by four parameters: $(N, k, \lambda, \mu)$. Each vertex is connected to $k$ other vertices through graph edges so all vertices have degree $k$. If two vertices are connected through a graph edge then the number of other vertices which are connected to both of them is $\lambda$. But if two vertices are not connected through a graph edge then this number is $\mu$. Note that the notation $k$ and $\lambda$ have different meanings in this subsection. A necessary condition that must be satisfied by these parameteres for the existence of a $SR(N)$ is 
\begin{equation}
	k(k - \lambda - 1) = (N - k - 1)\mu, \label{parametercondition}
\end{equation}
which also implies that $k$ is $\Omega(\sqrt{N})$. There are three eigenvalues of the Laplacian of a $SR(N)$. One is zero for which the corresponding eigenstate is $|s\rangle = \sum_{i}|i\rangle/\sqrt{N}$. Two other eigenvalues are given by 
\begin{equation}
F_{1} \pm F_{2},\ \ F_{1} = k - \frac{\lambda -\mu}{2},\ \ F_{2} =  \sqrt{k - \mu + \frac{(\lambda - \mu)^{2}}{4}}.
\end{equation}
We show that $F_{1} \gg F_{2}$ if the following condition is satisfied, i.e., 
\begin{equation}
N\mu \gg k - \mu + \frac{(\lambda - \mu)^{2}}{4} \Longrightarrow F_{1} \gg F_{2}. \label{parameterassumption}
\end{equation}
To show this, we use the fact that 
\begin{eqnarray}
F_{1}^{2} - \frac{(\lambda - \mu)^{2}}{4} & = & k^{2} + (\mu - \lambda) k \\ \nonumber
& = & k(k-\lambda-1) + k(\mu + 1) \\ \nonumber
& = & \mu(N-k-1) + k(\mu + 1) \\ \nonumber
& = & N\mu + (k-\mu),    
\end{eqnarray}
where we have used Eq. (\ref{parametercondition}) in going from second to third step. As $F_{2}^{2} = (k-\mu) + \frac{(\lambda - \mu)^{2}}{4}$, we find that $F_{1} \gg F_{2}$ if the assumption (\ref{parameterassumption}) is true. In that case, the non-zero eigenvalues of the Laplacian are $E_{z\neq 0} = F_{1}(1+o(1))$. We choose $m$ to be $1$ in our analysis so that $|\sigma\rangle$ is $|s\rangle$ and $\alpha$ is $1/\sqrt{N}$. Also, $M_{r}$ is $1/F_{1}^{r}(1+o(1))$ the assumption (\ref{recipeassume}) is satisfied as $0 \ll F_{1}/\sqrt{N} \ll F_{1}$ for large $N$. Putting $M_{1} = 1/F_{1}$ in the evolution Hamiltonian $H = M_{1}L - |t\rangle\langle t|$, we find that $|\phi(T)\rangle$ is almost $|t\rangle$ as $M_{1}/\sqrt{M_{2}}$ is $1-o(1)$. The evolution time is $O(\sqrt{N})$ which is the optimal performance. 

	Thus as long as the assumption (\ref{parameterassumption}) is satisfied, an optimal quantum search is possible on a strongly regular graph. This assumption is true for Paley graphs and the Latin square graphs, the class of strongly regular graphs studied in ~\cite{globalsymmetry}. For Paley graphs, the parameters $(N,k,\lambda,\mu)$ satisfy
\begin{equation} 
N = 4t + 1,\  k = 2t,\ \lambda = t - 1,\  \mu = t.
\end{equation}
For the Latin square graphs, they satisfy
\begin{equation}
N = t^2,\ k = d(t-1),\ \lambda = d^2 - 3d + t,\ \mu = d(d-1).
\end{equation}
Note that the notations $t$ and $d$ have different meanings in this subsection. This is easy to check that the assumption (\ref{parameterassumption}) is satisfied and $F_{1} \gg F_{2}$ for Paley graphs for $N \gg 1$ as then $t \gg 1$ and $F_{1} \approx 2t$ whereas $F_{2} \approx \sqrt{t}$. The assumption is also satisfied for Latin square graphs for $N \gg 1$ (which implies $t =\sqrt{N} \gg 1$) and for $1 \ll d \ll t$. Then $F_{1} \approx td$ whereas $F_{2} \approx t/2$. Thus it is possible to get optimal quantum search on Paley and Latin square graphs. In ~\cite{globalsymmetry}, authors have presented a different analysis for quantum search on strongly regular graphs using degenerate perturbation theory. Our analysis offers an alternative to this. 

	Latin square graphs are proved to be asymmetric for large $N$ unlike the complete graphs, cubical lattices or the hypercubes which exhibit a global symmetry. The optimality of quantum search on Latin square graphs was used in ~\cite{globalsymmetry} to argue that global symmetry is not necessary for optimal quantum search.

\section{Discussion and Conclusion}

	We have analysed the dynamics of quantum search on general graphs. We have found that the performance of quantum search is mainly determined by two parameters $M_{1}$ and $M_{2}$ of the graph. Thus we have developed the criteria any graph must satisfy to allow a successful quantum search.


\begin{thebibliography}{99}

\bibitem{grover1} L.K. Grover, Phys. Rev. Lett. \textbf{79}, 325 (1997).
\bibitem{grover2} L.K. Grover, Phys. Rev. Lett. \textbf{80}, 4329 (1998).
\bibitem{childs1} A. M. Childs and J. Goldstone, Physical Review A, 70:022314, 2004.
\bibitem{connectivity} D. A. Meyer and T. G. Wong, Physical Review Letters, 114:110503, 2015.
\bibitem{globalsymmetry} J. Janmark, D. A. Meyer, and T. G. Wong, Physical Review Letters, 112:210502, 2014.
\bibitem{regularity} L. Novo, S. Chakraborty, M. Mohseni, H. Neven, and Y. Omar, Scientific Reports, 5:13304, 2014.
\bibitem{erdosrenyi} S. Chakraborty, L. Novo, A. Ambainis, and Y. Omar, Phys. Rev. Lett. \textbf{116}, 100501 (2016). 
\bibitem{optimal}C. Bennett, E. Bernstein, G. Brassard, and U. Vazirani, SIAM J. Computing {\bf 26}, 1510 (1997) [arXiv.org:quant-ph/9701001].
\bibitem{simplexlattice} D. Dhar, J. Math. Phys. 18, 577 (1977).
\bibitem{weightedsimplex} T.G. Wong, Phys. Rev. A {\bf 92}, 032320, (2015).
\bibitem{hypercube1}E. Farhi, J. Goldstone, S. Gutmann, and M. Sipser, arXiv.org:quant-ph/0001106. 
\bibitem{hypercube2}A. Childs, E. Deotto, E. Farhi, J. Goldstone, S. Gutmann, and A. J. Landahl, Phys. Rev. A {\bf 66}, 032314 (2002).
\bibitem{erdos} P. Erd{\H{o}}s and A. R{\'{e}}nyi, Publ. Math. Debrecen, 6:290--297, (1959).
\bibitem{renyi}P. Erd\H{o}s and A. R\'{e}nyi, Publications of the Mathematical Institute of the Hungarian Academy of Sciences, 5:17--61, (1960).
\bibitem{erdoseigen} V. H. Vu, Spectral norm of random matrices,
Combinatorica, 27(6):721--736, (2007).
\end{thebibliography}
\end{document}